\documentclass[aps,preprint,superscriptaddress,amsmath,groupedaddress,showpacs,tightenlines,nofootinbib]{revtex4-1}
\usepackage{graphicx}  %
\usepackage{amsmath}
\usepackage{bm}  %
\usepackage{tikz}
\usepackage{epsfig}
\usepackage{times}
\usepackage{color}
\usepackage{hyperref}
\newcommand{\bea}{\begin{eqnarray}}
\newcommand{\eea}{\end{eqnarray}}
\newcommand{\beq}{\begin{equation}}
\newcommand{\eeq}{\end{equation}}
\newcommand{\bqa}{\begin{eqnarray}}
\newcommand{\eqa}{\end{eqnarray}}

\def\mqo2{{\!\!\!}}

\newcommand{\matrixel}[3]{\left< #1 \mvph{#2#3} \right| #2 \left| #3 \mvph{#1#2} \right>} % for Dirac matrix elements
\newcommand{\alignmentpoint}{&}
\newcommand{\mvph}[1]{\vphantom{\def\alignmentpoint{}#1}}
\newcommand{\abs}[1]{\lvert#1\rvert}

\def\fun#1#2{\lower3.6pt\vbox{\baselineskip0pt\lineskip.9pt
  \ialign{$\mathsurround=0pt#1\hfil##\hfil$\crcr#2\crcr\sim\crcr}}}
\begin{document}

\title{Tetramer Bound States in Heteronuclear Systems}
\author{C. H. Schmickler}
\email{schmickler@theorie.ikp.physik.tu-darmstadt.de}
\affiliation{Institut f\"ur Kernphysik, Technische Universit\"at Darmstadt,
64289\ Darmstadt, Germany}
\affiliation{RIKEN Nishina Center, RIKEN, Saitama 351-0198, Japan}
\author{H.-W. Hammer}
\affiliation{Institut f\"ur Kernphysik, Technische Universit\"at Darmstadt,
64289\ Darmstadt, Germany}
\affiliation{ExtreMe Matter Institute EMMI, GSI Helmholtzzentrum f\"ur 
Schwerionenforschung, 64291\ Darmstadt, Germany}
\author{E. Hiyama}
\affiliation{RIKEN Nishina Center, RIKEN, Saitama 351-0198, Japan}
\date{\today}
%\date{November 2007}

\begin{abstract}
We calculate the universal spectrum of trimer and tetramer states 
in heteronuclear mixtures of ultracold atoms with different masses in 
the vicinity of the heavy-light dimer threshold. To extract the energies, we 
solve the three- and four-body problem for simple two- and
three-body potentials tuned to the universal region
using the Gaussian expansion method.
We focus on the case of one light particle of mass $m$ and two or three 
heavy bosons of mass $M$ with resonant heavy-light interactions. 
We find that trimer and tetramer cross into the heavy-light dimer threshold 
at almost the same point and that as the mass ratio $M/m$ decreases,
the distance between the 
thresholds for trimer and tetramer states becomes smaller. 
We also comment on the possibility of observing exotic 
three-body states consisting of a dimer and two atoms in this
region and compare with previous work.
\end{abstract}

\smallskip
\maketitle

%%%%%%%%%%%%%%%%%%%%%%%%%%%%%%%%%%%%%%%%%%%%%%%
\section{Introduction}
\label{sec:intro}
%%%%%%%%%%%%%%%%%%%%%%%%%%%%%%%%%%%%%%%%%%%%%%%

\begin{figure}[htb]
\bigskip
\centerline{\includegraphics*[width=8cm,angle=0]{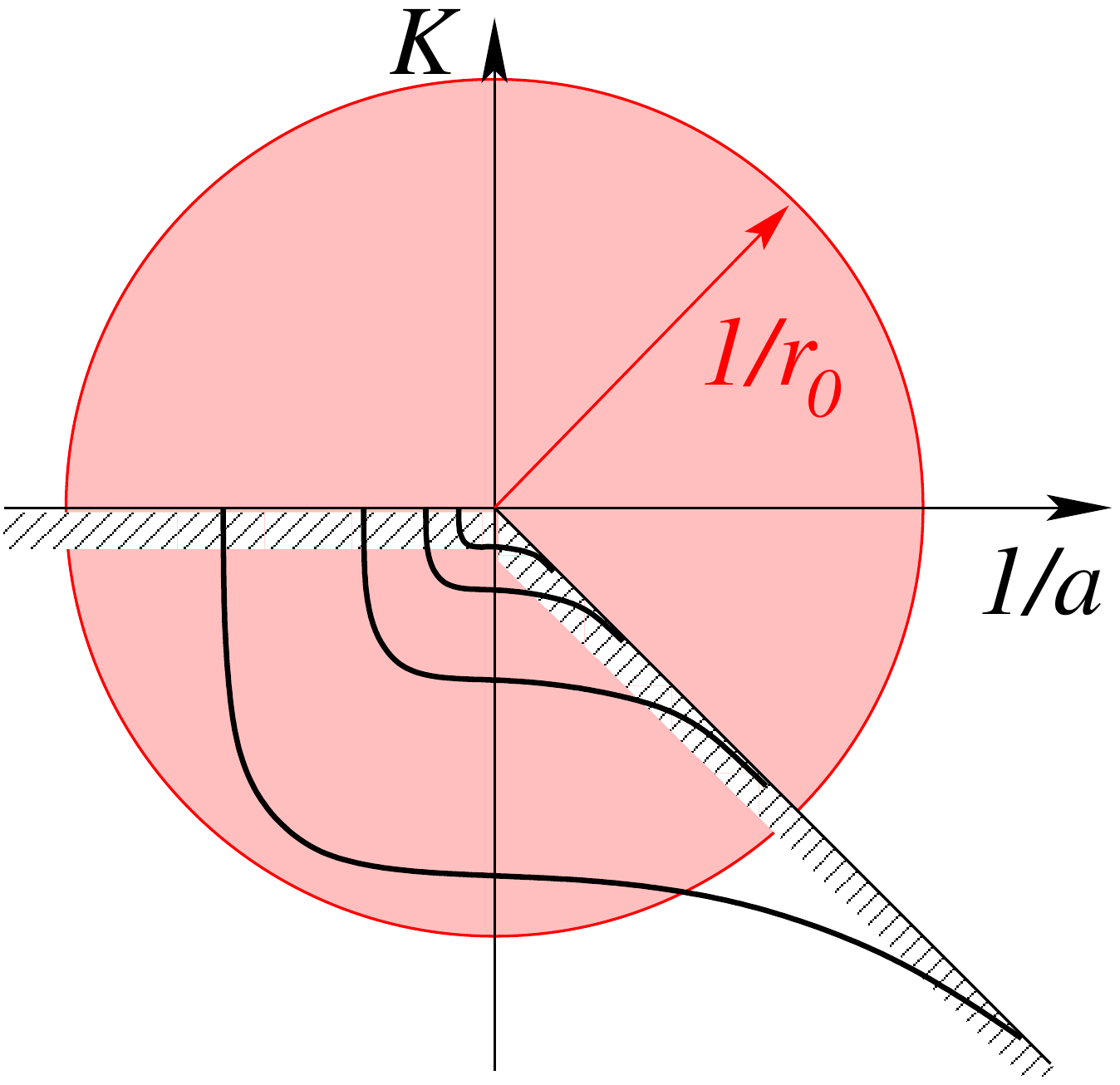}}
\medskip
\caption
{The $a^{-1}$--$K$ plane for the 3-body problem. 
Scattering thresholds are indicated by the cross-hatching while the
universal window is given by the circle of radius $1/r_0$.
A few of the infinitely many branches of Efimov states are shown
by the solid lines.}
\label{fig:3body}
\end{figure}

Particles with resonant short-range interactions  
display the Efimov effect \cite{Efimov70} and
related universal phenomena associated with a discrete scaling
symmetry \cite{Braaten:2004rn,Platter:2009gz,Naidon:2016dpf}.
In particular, Efimov showed that three identical bosons form
infinitely many trimer bound states with an accumulation point at the
scattering threshold when the $s$-wave scattering length $a$ is tuned
to the unitary limit $1/a =0$:
%----------------------
\begin{eqnarray}
E^{(n)}_3 = -(e^{-2\pi/s_0})^{n} \:\frac{\hbar^2 \kappa^2_*}{m},
\label{kappa-star}
\end{eqnarray}
%----------------------
where $m$ is the mass of the particles, $s_0 = 1.00624...$ is
a transcendental number, and
$\kappa_*$ is the binding wavenumber of the Efimov state
labeled by $n=0$.  The geometric  spectrum in (\ref{kappa-star})
is the signature of a
discrete scaling symmetry with scaling factor $e^{\pi/s_0}\approx 22.7$.
For a finite scattering length larger than the range of the
interaction, the universal properties persist but
there is only a finite number of Efimov states.
The full structure of the trimer spectrum as a function of
the inverse scattering length $1/a$ is conveniently summarized
in the so-called Efimov plot illustrated in Fig.~\ref{fig:3body}.
This plot shows the scattering length
dependence of the bound states in the 
%%%%%%%%%%%%%%%%%%%%%%%%%%%%%%%%%%%%%%%%%%%%%%%%
%%%%%%%%%%%%%%%%%%%%%%%%%%%%%%%%%%%%%%%%%%%%%%%%%
$a^{-1}$--$K$ plane, where $K={\rm sgn}(E)\sqrt{mE}$
is a momentum variable that corresponds to the binding momentum
for bound states with $E<0$.
The threshold for scattering states is indicated by the hatched area.
The Efimov trimers are represented by the solid
lines below the threshold. 
There are infinitely many branches of Efimov trimers, 
but only a few are shown. In the unitary limit $1/a=0$, infinitely
many trimer states accumulate at threshold. 
When the radial variable $\sqrt{K^2 +1/a^2}$ is of the order of the
inverse interaction range $1/r_0$ or larger, universality breaks down
and the states become sensitive to
the details of the interaction at short distances.

Ultracold atoms are an ideal tool to study such phenomena since
the scattering length can be tuned experimentally 
to be large compared to the range of the interaction using Feshbach resonances. 
Efimov trimers can be observed in ultracold atomic gases 
via their signature in three-body recombination rates
\cite{NM-99,EGB-99,BBH-00,BH01,Braaten:2003yc}. Using this method,
Kraemer {\it et al.}\ provided the first evidence for Efimov trimers 
in an ultracold gas of $^{133}$Cs atoms by observing the resonant 
enhancement of atom losses through three-body recombination 
caused by the trimers \cite{Kraemer:2006}. Efimov states have been
observed in a variety of other atomic species since then
\cite{Ferlaino:2010,Naidon:2016dpf}.

The scaling factor  $e^{\pi/s_0}$ can be significantly reduced in heteronuclear systems 
with different mass particles which makes it easier to observe multiple states.
In this paper, we focus on heteronuclear systems 
with two species of atoms where only the interspecies scattering length 
is large.  
For comparable masses the scaling factor is quite large 
in this case ($e^{\pi/s_0}\approx 1986.1$ for equal masses) as we now have 
only two resonant interactions out of three. However, in the case of
two heavy bosonic atoms of mass $M$ (denoted $H$) and one light atom
of mass $m < M$ (denoted $L$), this factor can become 
significantly smaller than the value 22.7 for identical 
bosons~\cite{Amado72,Efimov72,Efimov73,Braaten:2004rn}.  
Relaxation and recombination losses near an interspecies resonance have 
recently been investigated in mixtures of Cesium and Lithium atoms
and three consecutive Cs-Cs-Li Efimov resonances  were observed
in agreement with the predicted scaling relations
\cite{Tung:2014,Johansen:2016,Pires:2014,Ulmanis:2015,Ulmanis:2016}.  The effect
of the Cs-Cs scattering length on the Efimov resonances was investigated in
Refs.~\cite{Ulmanis:2016hdd,Haefner:2017,Acharya:2016kjr}.\footnote{Note
  that there should also be Li-Li-Cs Efimov resonances, but in this
case the scaling factor is extremely large.}

Connected to the Efimov states are universal bound states in the 
four- and higher-body sector.
Although there is no accumulation of bound states at threshold, there
are universal four- and higher-body states attached to each Efimov
trimer. The binding energies of these higher-body states are
uniquely determined by $a$ and $\kappa_*$ and no higher-body
parameters are required \cite{Platter:2004qn}.
In experiments with ultracold Cesium atoms
universal states up to $N=5$ have been observed \cite{zenesinihuang2013}.
For identical bosons, exactly two tetramer
states are attached to each trimer state
\cite{Hammer:2006ct,vStech08,Deltuva:2010xd,Deltuva:2012ig}.
This pattern appears
to hold also for the higher-body states and has been calculated explicitly
up to $N=6$ \cite{vStech09,Gattobigio:2012tk}. The ground states
have been calculated up to
$N=16$ \cite{Nicholson:2012zp,Kievsky:2014dua,Blume:2015}.

%inserted part
The demonstration of the existence of the higher-body universal states in 
experiments with mixtures represents a frontier
  in the physics of ultracold atoms.
Only recently a loss feature attributed to a Na$_2$Rb$_2$ state has 
been seen in an experiment involving NaRb Feshbach molecules \cite{wangye2016}.
%end of inserted part
Focusing on four-body
states, the $H_3 L$ states are very promising due to
the favorable scaling factor. However, the relation of the higher-body
energies to the trimer energies is not governed by the same universal numbers
and even the number of higher-body states attached to the trimers will
very likely depend on the mass ratio.
The signatures of Efimov physics in heteronuclear four-body systems were
first investigated in Ref.~\cite{wanglaing2012}.
A detailed study of the four-body states in a mixture
of Cesium and Lithium atoms in the vicinity of the unitary
limit and the three-atom threshold
was recently performed by Blume and Yan \cite{Blume2014}.
For large enough mass ratios ($M/m \geq 16$) they found two universal four-body states in the unitary limit. 
They also found an excited four-body state for negative values of the scattering length for all mass 
ratios they investigated ($8 \leq M/m \leq 50$). For the mass ratio $M/m = 133/6$ they showed that 
the excited tetramer becomes unbound at a certain positive value of the scattering length, leading them 
to the conjecture that the scattering length at which the excited tetramer becomes unbound depends 
on the mass ratio.

In this paper, we build upon the investigation by Blume and Yan and
extend it in several ways. We extend their calculation to positive
scattering lengths further away from the unitary limit. In particular, we
investigate the mass dependence of the spectrum and discuss the number
of universal tetramer states as function of the mass ratio. 
One of our key findings is that the structure of the spectrum changes as 
the mass ratio $M/m$ decreases. In particular, the distance between the 
thresholds for trimer and tetramer states gets smaller 
until they almost coincide for $M/m\approx 1$. Preliminary studies of this aspect were presented in
Ref.~\cite{Schmickler:2016iif}.
Moreover,
we summarize our insights in a multi-dimensional Efimov plot which shows
$K$ as a function of $1/a$ and the ratio of the heavy and light masses
$M/m$. 

A particularly interesting aspect of the $H_3 L$ system is the prospect of
observing effective Efimov states of a $HL$ dimer
and two heavy atoms $H$ which are denoted $H_2(HL)$ in the following.
Such an effective
Efimov effect must exist close to the crossing point of the
$H_2 L$ trimer and the $HL$ dimer threshold.
This requires precise studies of the
bound state spectrum near the $HL$ dimer threshold. 
Assuming that only the $HL$ dimer and the two $H$ atoms interact, one would
expect a scaling factor that is larger than 1986, which is the limiting
value for $M/m\to \infty$~\cite{Braaten:2004rn}.
For a Cesium-Lithium mixture, one would expect a scaling factor of
about 2500.

Using the Born-Oppenheimer approximation, Wang and
collaborators~\cite{wanglaing2012}  
found evidence for the existence of two such states for the
mass ratio $M/m=50$
with a scaling factor of approximately 20. They explain the small
scaling factor with an effective interaction between the $H$ particles
that is induced by the exchange of the $L$ particle. 
In this paper, we will address this puzzle by investigating
the effective $H_2(HL)$ Efimov states for different mass
ratios $M/m$ using explicit four-body
calculations as well as using an effective field theory for heteronuclear
three-body mixtures \cite{Helfrich:2010yr}.

%%%%%%%%%%%%%%%%%%%%%%%%%%%%%%%%%%%%%%%%%%%%%%%
\section{Method}
% \label{sec:interaction}
%%%%%%%%%%%%%%%%%%%%%%%%%%%%%%%%%%%%%%%%%%%%%%%

To obtain the energies of  the $H_2L$ trimer and $H_3 L$ tetramer systems,
we employ the Gaussian expansion method (GEM) \cite{hiyamakino2003}. 
The method was successfully applied to
various types of three- , four- and five-body
hypernuclear systems \cite{Hiyama96,Hiyama97,
Hiyama99,Hiyama01,Hiyama02,Hiyama10} and
trimer and tetramer systems of $^4$He \cite{Hiyama-atom2012a,
Hiyama-atom2012b,Hiyama-atom2014}. 

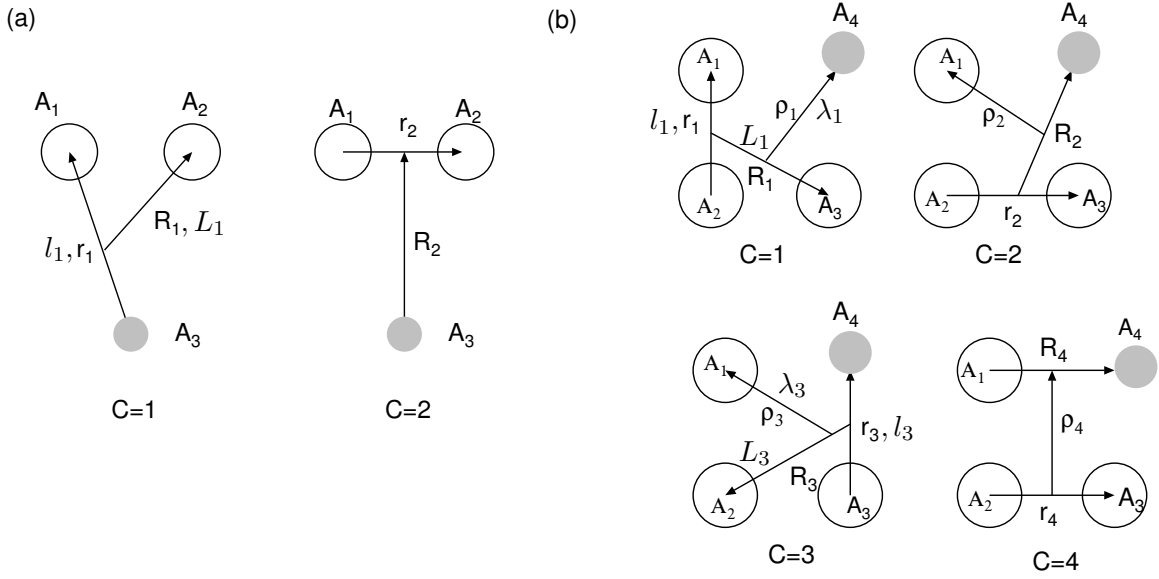
\begin{figure}[htb]
\bigskip
% \centerline{\raisebox{0.4cm}{\parbox{6cm}{\includegraphics*[height=4.8cm,angle=0]{atom-three.pdf}}}
% \quad\raisebox{0.cm}{\parbox{6cm}{\includegraphics*[height=6cm,angle=0]{atom-four.pdf}}}
% }
% \includegraphics[width=0.9\textwidth, trim={0 20em 0 0},clip]{fig2withlabels.pdf}
\begin{tikzpicture}
 \node at (0,0.99) {\includegraphics[width=0.4\textwidth]{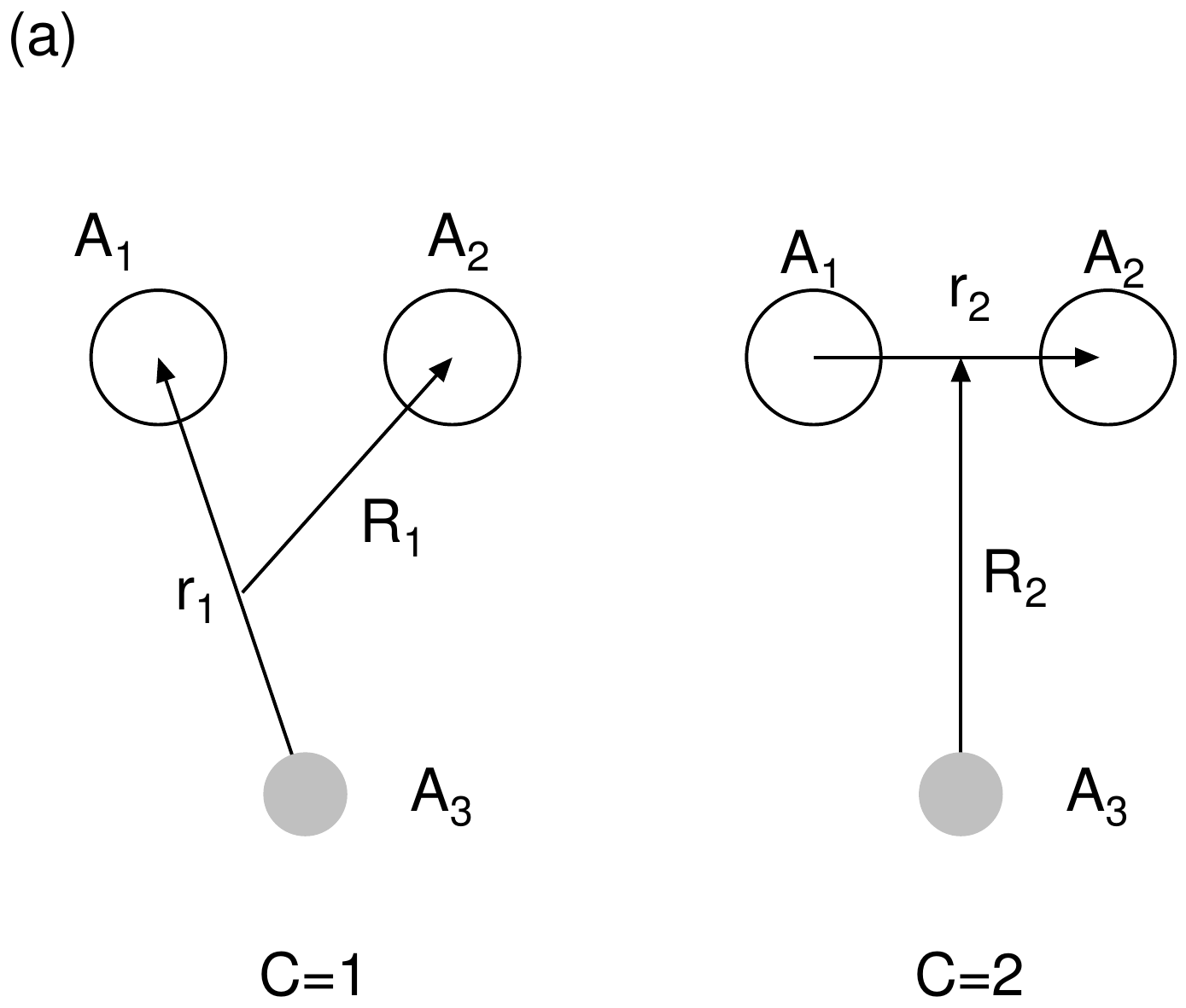}};
 \node at (8,0) {\includegraphics[width=0.5\textwidth]{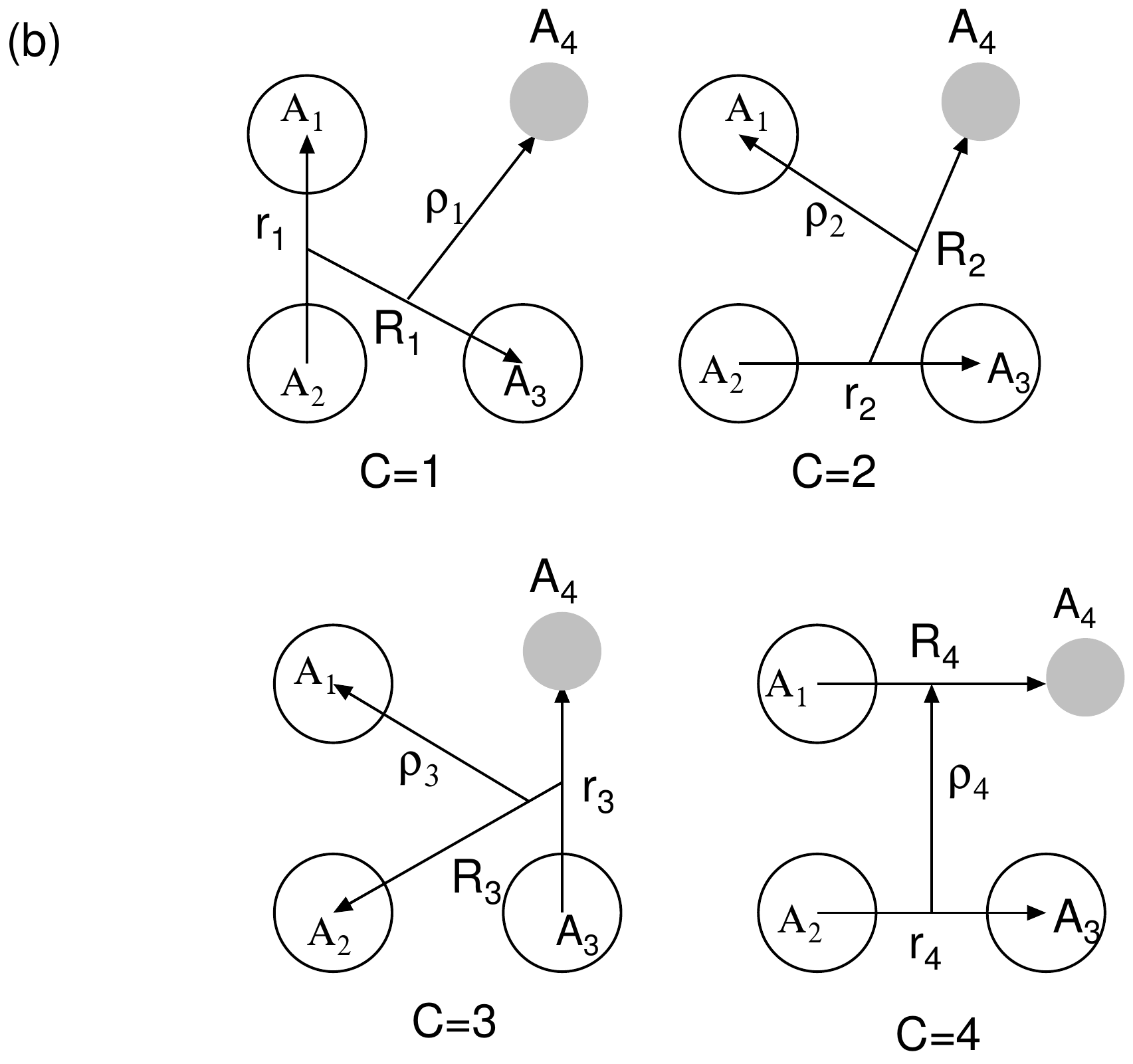}};
 \node at (-2.55,0.5) {$l_1,$};
 \node at (-0.63,0.85) {$, L_1$};
 \node at (5.5,2.2) {$l_1,$};
 \node at (6.7,2.) {$L_1$};
 \node at (7.7,2.35) {$\lambda_1$};
 \node at (8.6,-1.88) {$, l_3$};
 \node at (6.7,-2.2) {$L_3$};
 \node at (7.2,-1.3) {$\lambda_3$};
\end{tikzpicture}

\medskip
\caption{Jacobian coordinates for trimer (a) and tetramer (b) systems.
In (a) $A_1$ and $A_2$ are heavy bosonic atoms and $A_3$ is a light 
atom. In (b) $A_1$, $A_2$, and $A_3$ are heavy bosonic atoms 
and $A_4$ is a light atom. The two bosonic atoms in the trimer and the 
three bosonic atoms in the tetramer are symmetrized.}
\label{fig:three-atom}
\end{figure}

We use Jacobian coordinate sets for trimer and tetramer systems  as shown in Fig. \ref{fig:three-atom}.
The Hamiltonians for the three and four-body systems are described as
\begin{eqnarray}
H& = & T
+ \sum^2_{i=1} V_{A_i A_3} +V_{A_1 A_2 A_3} \nonumber \\
H &= &  T
+ \sum^3_{i=1} V_{A_i A_4} + \sum^3_{i<j} V_{A_i A_j A_4}
\label{eq:pot}
\end{eqnarray}
respectively, where $T$ is the kinetic-energy operator,
the $V_{A_i A_3}$ and $V_{A_i A_4}$ are two-body potentials and 
$V_{A_1 A_2 A_3}$ and  $V_{A_i A_j A_4}$ are three-body potentials, which are 
the same  as in Ref.~\cite{Blume2014}.
The details are  given below.
Here it should be noted that we switch off the two-body interaction between
the heavy bosonic atoms.

The energies are obtained by solving the
trimer and tetramer Schr\"{o}dinger equations
given by 
\begin{eqnarray}
(H-E)\Psi_{\rm trimer} &= &0, \nonumber \\
(H-E)\Psi_{\rm tetramer} &=&0.
\label{Hamil}
\end{eqnarray}

%%%%%%%%%%%%%%%%%%%%%%%%%%%%%%%%%%%%%%%%%%%%%%%%
%%%%%%%%%%%%%%%%%%%%%%%%%%%%%%%%%%%%%%%%%%%%%%%%%

The wave functions of three-body and four-body systems are
described as a sum of amplitudes for the rearrangement channels
($c=1,2$ for the trimer,
$c=1, ..., 4$ for the tetramer) illustrated in Fig.~\ref{fig:three-atom}
as follows:
\begin{equation} \label{wave}
 \begin{split}
  \Psi^{\rm trimer}_{JM} & =  \sum^2_{c=1} \sum_{n_c,N_c}\sum_{\ell_c,L_c} C^{(c)}_{n_c \ell_c N_c L_c} \\ 
&\quad \times [\phi^{(c)}_{n_c \ell_c}({\bf r}_c) \psi^{(c)}_{N_c L_c}({\bf R}_c)]_{JM} \\
\Psi^{\rm tetramer}_{JM} & =  \sum^4_{c=1} \sum_{n_c,N_c,\nu_c}\sum_{\ell_c,L_c,\lambda_c}  
C^{(c)}_{n_c \ell_c  N_c L_c \nu_c \lambda_c } \\ 
&\quad \times [[\phi^{(c)}_{n_c \ell_c}({\bf r}_c) \psi^{(c)}_{N_c L_c}({\bf R}_c)]_I \xi_{\nu_c \lambda_c}^{(c)}
(\mbox{\boldmath $\rho$}_c)]_{JM}.
 \end{split}
\end{equation}

Here, the two (three) bosonic atoms in the trimer (tetramer) must be
symmetrized.
For this purpose, we take the angular momenta $\ell $ to be even numbers in the 
corresponding rearrangement channels.
We take the functional form of $\phi_{n\ell m}({\bf r})$, $\psi_{NLM}({\bf R})$,
and $\xi_{\nu \lambda \mu}(\mbox{\boldmath $\rho$})$ as
\begin{align}
\phi_{nlm}({\bf r}) & =  \phi^{\rm G}_{nl}(r)\:Y_{lm}({\widehat {\bf r}}) ,&
 \phi^{\rm G}_{nl}(r) &=  N_{nl}\,r^l\:e^{-(r/r_n)^2},
\label{eq:3gaussa}\\
\psi_{NLM}({\bf R}) & =  \psi^{\rm G}_{NL}(R)\:Y_{LM}({\widehat {\bf R}}) , &
 \psi^{\rm G}_{NL}(R) & = 
N_{NL}\,R^L\:e^{-(R/R_N)^2},
\label{eq:3gauss} \\
\xi_{\nu \lambda \mu}(\mbox{\boldmath $\rho$}) & =  
 \xi^{\rm G}_{\nu \lambda \mu}(\mbox{\boldmath $\rho$})\:Y_{\lambda \mu}
({\widehat {\mbox{\boldmath $\rho$}}}) , &
 \xi^{\rm G}_{\nu \lambda}(\rho) & = 
N_{\nu \lambda}\,\rho^\lambda\:e^{-(\rho/\rho_\nu)^2},
\label{eq:3gaussb}
\end{align}
where
$N_{nl}$, $N_{NL}$, $N_{\nu \lambda}$  denote the normalization constants.
The Gaussian range parameters are chosen according to
geometrical progression:
\begin{align}
 r_n& =r_{\rm min}\, a^{n-1},&                   a& =\left(\frac{r_{\mathrm{max}}}{r_{\mathrm{min}}}      \right)^{\frac{1}{n_{\mathrm{max}}-1}}& (n& =1 \sim n_{\rm max})\;, \\
 R_n& =R_{\rm min}\, A^{N-1},&                   A& =\left(\frac{R_{\mathrm{max}}}{R_{\mathrm{min}}}      \right)^{\frac{1}{N_{\mathrm{max}}-1}}& (N& =1 \sim N_{\rm max})\;, \\
 \rho_n& =\rho_{\rm min}\, \alpha^{\nu-1},& \alpha& =\left(\frac{\rho_{\mathrm{max}}}{\rho_{\mathrm{min}}}\right)^{\frac{1}{\nu_{\mathrm{max}}-1}}& (\nu& =1 \sim \nu_{\rm max})\;.
\label{eq:progR}
\end{align}

The Gaussian ranges are very suitable for accurately
describing both the short-range correlations and
the long-range tail in the asymptotic region of few-body wave functions.
The Gaussian shape of basis functions makes
the calculation of matrix elements simple even between basis functions of 
different channels.

%%%%%%%%%%%%%%%%%%%%%%%%%%%%%%%%%%%%%%%%%%%%%%%%%%%%%%%%%%%%
\begin{table}[htb]
\caption{Typical three-body angular momentum space
and Gaussian range parameters
employed
for the trimer system. The channel numbers refer to 
Fig.~\ref{fig:three-atom}.
}
\label{table:parameters}
\begin{center}
\begin{tabular}{rrrrrrrrrrrrr}
\hline 
\hline 
\vspace{-3 mm} \\
 $c$& $l_c$& $L_c$& \hspace{2em}$n_{\rm max}$& \hspace{1em}$r_{\rm min}$& \hspace{1em}$r_{\rm max}$& \hspace{2em}$N_{\rm max}$& \hspace{1em}$R_{\rm min}$& \hspace{1em}$R_{\rm max}$\\
\hline 
1& 0& 0&        23&     0.028&  15&     20&     0.038&  65\\
1& 2& 2&        21&     0.180&  27&     19&     0.228&  20\\
2& 0& 0&        19&     0.110&  26&     21&     0.085&  29\\
\vspace{-3 mm} \\
\hline 
\hline \\
\end{tabular}
\end{center}
\end{table}
%%%%%%%%%%%%%%%%%%%%%%%%%%%%%%%%%%%%%%%%%%%%%%%%%%%%%%%%%%%%%

The eigenenergies $E$ in Eq.~(\ref{Hamil}) and
the coefficients $C$ in Eq.~(\ref{wave}) are then determined by
the Raleigh-Ritz variational method.

In Tables \ref{table:parameters} and \ref{table:parameters2},
typical parameters of the basis functions for the trimer and
tetramer are listed.

%%%%%%%%%%%%%%%%%%%%%%%%%%%%%%%%%%%%%%%%%%%%%%%%%%%%%%%%%%%%%%%%%
\begin{table}[htb]
\caption{Typical four-body angular momentum space 
and Gaussian range parameters employed for the tetramer system.
The channel numbers refer to 
Fig.~\ref{fig:three-atom}.}
\label{table:parameters2}
\begin{center}
\hspace{-4em}\begin{tabular}{rrrrrrrrrrrrrrrrr}
\hline 
\hline 
\vspace{-3 mm} \\
  $c$  &$l_c$   &$L_c$  & \hspace{1em}$n_{\rm max}$  &\hspace{1em}$r_{\rm min}$ &\hspace{1em}$r_{\rm max}$ & \hspace{2em}$N_{\rm max}$  &\hspace{1em}$R_{\rm min}$ &\hspace{1em}$R_{\rm max}$ 
  &\hspace{2em}$\nu_{\rm max}$  &\hspace{1em}$\rho_{\rm min}$ &\hspace{1em}$\rho_{\rm max}$\\
\hline 
1& 0& 0&        11&     0.073&  13&     13&     0.059&  8&      11&     0.055&  14\\
2& 0& 0&        15&     0.043&  9&      15&     0.044&  7&      15&     0.075&  33\\
3& 0& 0&        17&     0.038&  6&      16&     0.044&  13&     17&     0.055&  25\\
3& 1& 1&        13&     0.057&  7&      13&     0.052&  6&      14&     0.094&  25\\
4& 0& 0&        14&     0.139&  15&     12&     0.033&  5&      14&     0.127&  12\\
\vspace{-3 mm} \\
\hline 
\hline \\
\end{tabular}
\end{center}
\end{table}
%%%%%%%%%%%%%%%%%%%%%%%%%%%%%%%%%%%%%%%%%%%%%%%%%%%%%%%%%%%%%%%%%%%%%

The potential and kinetic energy terms are expanded in the truncated basis as
shown in Eq.~(\ref{wave}). 
Because the states are not orthogonal this transforms the 
Schr\"odinger equation  
\begin{equation}
 \sum_{c}\sum_{n_cl_cN_cL_c}\matrixel{[\phi^{(c')}_{n_{c'} \ell_{c'}}({\bf r}_{c'}) \psi^{(c')}_{N_{c'} L_{c'}}({\bf R}_{c'})]_{JM}}{H - 
 E}{%
 C^{(c)}_{n_cl_cN_cL_c}[\phi^{(c)}_{n_c \ell_c}({\bf r}_c) \psi^{(c)}_{N_c L_c}({\bf R}_c)]_{JM}
} = 0
\end{equation}
into a general eigenvalue problem
\begin{equation}
 \sum_{\tilde n}(H_{\tilde n'\tilde n} - EN_{\tilde n'\tilde n})C_{\tilde n} = 0,
\end{equation}
where $N_{\tilde n'\tilde n}$ is a normalization matrix and $\tilde n$ is a
shorthand notation for all the indices that are summed over,
i.e. $\tilde n \stackrel{\wedge}{=} (c,n_c,l_c,N_c,L_c)$ in the three-body case. The matrix elements can be calculated analytically 
for some potentials and the equation can be solved using standard 
linear algebra methods.

%inserted part
Care has to be taken to avoid overcompleteness of the basis sets. We meticulously checked the convergence 
behavior and the behavior of the binding energies under slight variations of the potential depth to 
discard overcomplete basis sets.
%end of inserted part
%%%%%%%%%%%%%%%%%%%%%%%%%%%%%%%%%%%%%%%%%%%%%%%%%%%%%%%%%%%%%%%%%%%%%%%%%%%%%%%%%%%%%%%%%%%%%%%%%%%

%%%%%%%%%%%%%%%%%%%%%%%%%%%%%%%%%%%%%%%%%%%%%%%
\section{Interaction}
\label{sec:interaction}
%%%%%%%%%%%%%%%%%%%%%%%%%%%%%%%%%%%%%%%%%%%%%%%
Since we want to study universal effects, i.e. effects that are independent 
of the exact shape of the potential and only depend on the long-range behavior
of the potential, it is beneficial to use simple potentials.
We have chosen Gaussian potentials, because their matrix elements can be
calculated analytically in the Gaussian Expansion 
Method (GEM). There are many prior 
calculations with Gaussian potentials \cite{gattobigiokievsky11,Blume2014},
so we can easily validate our method.

We use the same potentials as in Ref.~\cite{Blume2014}.
The two-body potential is given by
\begin{equation}
 V_{A_i A_N} = v_0e^{-\frac{r_{iN}^2}{2r_0^2}},
 \label{eq:2bpot}
\end{equation}
and the three-body potential is 
\begin{equation}
 V_{A_i A_j A_N} = w_0e^{-\frac{r_{ij}^2 + r_{iN}^2 + r_{jN}^2}{16r_0^2}},
 \label{eq:3bpot}
\end{equation}
where $r_{ij} = r_i - r_j$, and $r_i$ is the position of atom $i$.
The interaction takes place only between different atoms, where the $N$th
atom is the special one (type $L$), 
and all other atoms are identical bosons (type $H$). 
The interaction between identical bosons is neglected,
because it is assumed to be very weak compared to the resonant
$HL$ interaction. This is reflected in the summation in Eq.~(\ref{eq:pot}). 

Gaussian potentials that are tuned to the universal regime
can be understood as regularized contact terms in an EFT expansion
\cite{gattobigiokievsky11}, so our
results are valid only in the energy regime of this implicit EFT expansion.
This requires that all calculated energies are small compared to the 
natural energy scale set by the window of universality
shown in Fig.~\ref{fig:3body}.
The natural energy scale connected to the potential (\ref{eq:2bpot},
\ref{eq:3bpot}) is
\begin{equation}
  E_s = \frac{\hbar^2}{2\mu r_0^2}\,,\qquad\mbox{ with }\quad
  \mu = \frac{mM}{m+M}
\label{eq:E_s}  
\end{equation}
the reduced mass of the interacting dimer ($HL$).
The two-body potential (\ref{eq:2bpot})
is attractive, whereas the three-body potential  (\ref{eq:3bpot})
is chosen to be repulsive such that the calculated binding energies
are in the window of universality, i.e. $B \ll E_s$. 
The three-body force parameter, $w_0$, is kept constant throughout the
calculations presented here, except where noted otherwise.

The units are chosen so that the calculations are numerically well-behaved,
and the results are presented in 
terms of $E_s$ and $a$, the $HL$ s-wave scattering length. 
We vary $v_0$ to reproduce different scattering lengths $a$ as $v_0$ is 
roughly linearly proportional to $1/a$. The length scale $r_0$ is then
essentially a free parameter that sets the scale for $a$.
Its numerical value is set to $0.04$ throughout our calculations.

%%%%%%%%%%%%%%%%%%%%%%%%%%%%%%%%%%%%%%%%%%%%%%%

\section{Benchmarking}
\begin{table}
\begin{tabular}{lllll}
$w_0/E_s$&\multicolumn{2}{c}{$\sqrt{\vphantom{E^{0,0}}E_3^0/E_s}$}&\multicolumn{2}{c}{$\sqrt{\vphantom{E^{0,0}}E_3^0/E_3^1}$}\\
& our result & \cite{Blume2014} & our result & \cite{Blume2014}\\
\hline
0.00    & 0.314347\hphantom{3ex}       & 0.314348\hphantom{3ex}      & 3.9341\hphantom{3ex}        & 3.9326\hphantom{3ex}        \\ 
0.16    & 0.188463       & 0.188467      & 4.1225        & 4.1205        \\ 
2.56    & 0.031410       & 0.031417      & 4.8955        & 4.8968        \\ 
9.60    & 0.025999       & 0.026004      & 4.9023        & 4.9025        \\ 

\end{tabular}
\caption{Comparison of our benchmarking results with \cite{Blume2014}
for the mass ratio $M/m = 133/6$. The numbers are taken from Fig. 2 of the
supplement and were provided to us by D. Blume \cite{Blume-private}.
The first column gives the strength of the three-body force in units
of the natural energy scale $E_s$ as defined in Eq.~\eqref{eq:E_s}. The following 
columns give the ratios of the trimer ground state and $E_s$
and the ratio of the lowest two trimer states, respectively.} 
\label{tab:trimerE}
\end{table}
%%%%%%%%%%%%%%%%%%%%%%%%%%%%%%%%%%%%%%%%%%%%%%%

%%%%%%%%%%%%%%%%%%%%%%%%%%%%%%%%%%%%%%%%%%%%%%%
\begin{table}
\begin{tabular}{lllll}
$w_0/E_s$&\multicolumn{2}{c}{$\sqrt{E_4^{0,0}/E_3^0}$}&\multicolumn{2}{c}{$\sqrt{E_4^{0,1}/E_3^0}$}\\
& our result & \cite{Blume2014} & our result & \cite{Blume2014}\\
\hline
0.00     & 1.89891\hphantom{3ex}& 1.89890\hphantom{3ex} & &   \\ 
2.56     & 1.51452       	& 1.51425       	& 1.0119\hphantom{3ex}  & 1.0116\hphantom{3ex}   \\ 
9.60     & 1.51121       	& 1.51119       	& 1.0110        	& 1.0106        	\\ 
\hline
\end{tabular}
\caption{Comparison of our benchmarking results with \cite{Blume2014}
for the mass ratio $M/m = 133/6$. The numbers are taken from Fig. 3 of the
supplement and were provided to us by D. Blume \cite{Blume-private}. 
The first column gives the strength of the three-body force in units
of the natural energy scale $E_s$ as defined in Eq.~\eqref{eq:E_s}.
The following columns show the ratio of the tetramer ground state and 
the trimer ground state and the ratio of the tetramer excited state and 
the trimer ground state, respectively.}
\label{tab:tetramerE}
\end{table}
%%%%%%%%%%%%%%%%%%%%%%%%%%%%%%%%%%%%%%%%%%%%%%%%%%%%%

In order to benchmark our method, we calculate the trimer
and tetramer energies in the vicinity of the unitary limit and compare 
with the results for the mass ratio $M/m = 133/6$
obtained by Blume and Yan \cite{Blume2014}. 
Our calculations are carried out very close to the unitary
limit at $\abs{a}/r_0 = 10^{10}$.  %genauer: -1.398e10
The corresponding results for the trimer energies are given in 
Table~\ref{tab:trimerE},
while the corresponding tetramer energies are given in 
Table~\ref{tab:tetramerE}.
For the different strengths of the three-body force $w_0$, 
we find agreement with Ref.~\cite{Blume2014} to three digits or better, 
which gives us confidence in our method. 

However, the time  spent optimizing the basis function parameters for the
actual calculations discussed in Sec.~\ref{4and3closetothresh}
was about $20$ times longer than for the benchmark calculations
discussed above. Moreover,
the basis size used for the actual calculations was about 
twice as large as for the benchmark calculations. This ensures
that our results in Sec.~\ref{4and3closetothresh} are converged 
to about 8 digits corresponding to a precision of $10^{-8} \sqrt{E_s}$.  

The energies are less sensitive to non-universal aspects of the
finite-range model potentials for larger values of $w_0$~\cite{Blume2014}.
As a consequence, we use $w_0/E_s\approx 9.6$ in our calculations 
of the $H_2 L$ and $H_3 L$ systems below. This ensures
absolute energies are small enough that range effects do not play a
role and the results are close to the universal limit of zero-range
interactions. In practice, we have ensured that all calculated energies satisfy 
$E \leq 0.13 E_s$.

Since the Gaussian Expansion Method as employed here uses a truncated sum over rearrangement channels and relative angular momenta to 
represent the wave function (cf.~Eq.~(\ref{wave})), contributions from higher relative angular momenta 
are neglected. We included as many configurations as possible without encountering numerical instabilities.  In addition, 
the states are extremely shallow in the region where very high precision
is required, such that higher relative angular momenta 
are strongly suppressed. The highest angular momentum quantum numbers that are included in our 
four-body calculations are $L_3=1, \lambda_3=1$. Including the corresponding 
$L_3=1, \lambda_3=1$ basis functions of the rearrangement channel $3$ (cf.~Fig.~\ref{fig:three-atom}), 
which contributes the most, 
leads to a correction of only $10^{-7} \sqrt{E_s}$ near the threshold.
This gives us confidence that including even higher relative angular momenta would not lead to 
significant contributions. 

\section{Efimov Plot for Heteronuclear Systems}
\begin{figure}[htb]
\bigskip
\centerline{\includegraphics*[width=12cm]{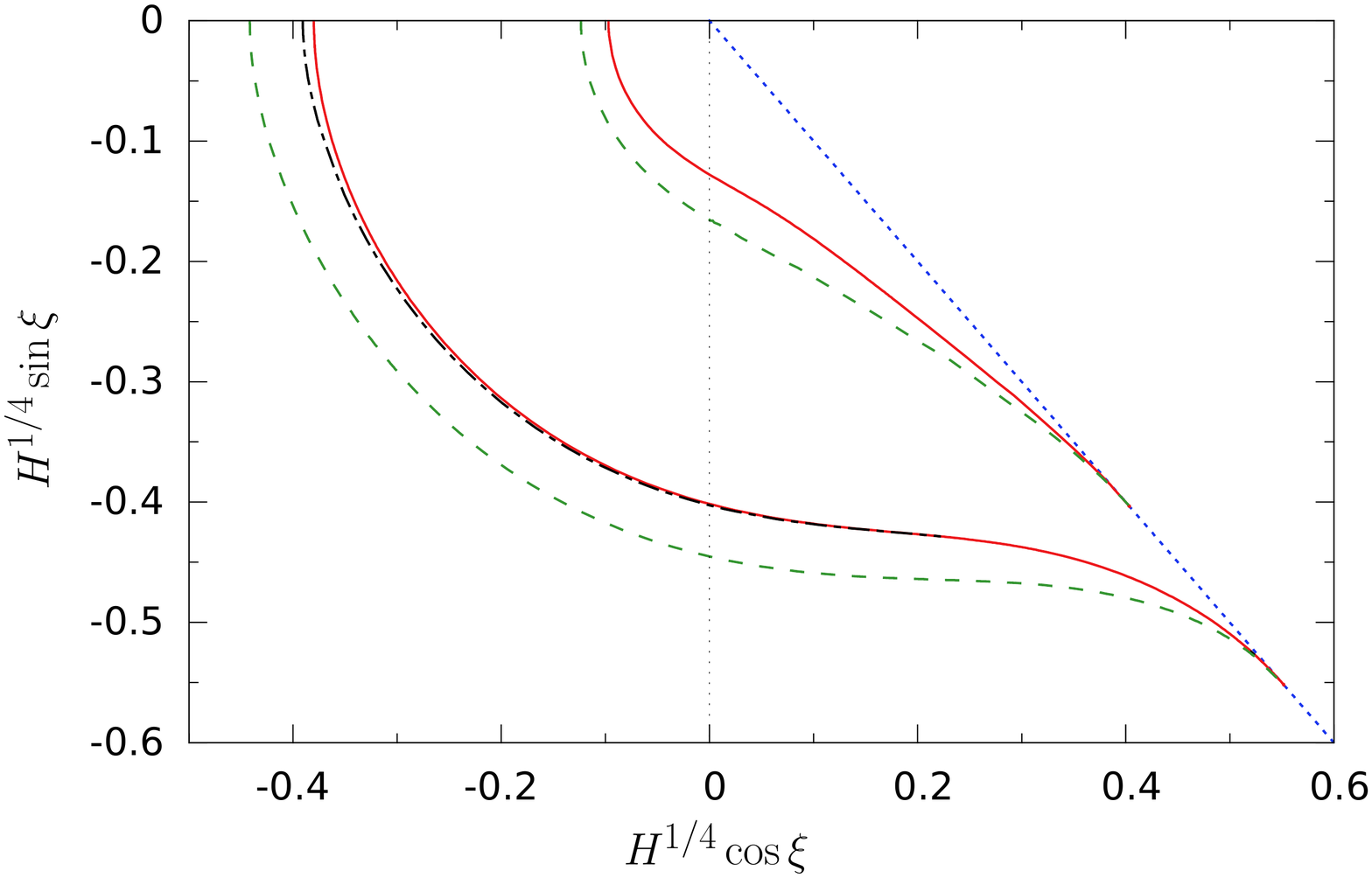}}
%\medskip
\caption{
  Efimov plot for $H_2 L$ trimers and $H_3 L$ tetramers in  a $^{133}$Cs-$^{6}$Li mixture (lower group of lines)
  and a $^7$Li-$^6$Li mixture (upper group of lines).
  The blue dotted line marks the $HL$ dimer + two $H$ atom threshold,
  the red solid lines the $H_2 L$ trimer + $H$ atom threshold, the dashed green 
  lines the $H_3 L$ tetramer ground states, and the dot-dashed black line the excited state of the tetramer. 
  To increase visibility of the features of the plot,
  the axes are rescaled by taking the fourth root. To this end, $H$ and $\xi$ are introduced 
  as $r_0/a = H\cos\xi$ and  $-\sqrt{\abs{E}/E_s} = H\sin\xi$  (cf.~Ref.~\cite{Hammer:2006ct}). 
}
\label{fig:overview}
\end{figure}

In this work, we focus on 
three-body systems of one light particle $L$ of mass $m$ and two heavy bosons
$H$ of mass $M$ and four-body systems of three heavy bosons $H$ and one
light particle $L$. 
In order to establish the 
qualitative behavior of trimer and tetramer states as a function 
of the mass ratio $M/m$, we have calculated the tetramer and trimer 
energies for the cases $(M=133,\,m=6)$, $(M=87,\,m=7)$, and
$(M=7,\,m=6)$, which correspond to possible mixtures of ultracold atoms,
as a function of $r_0/a$ where $r_0$ is the interaction
range of the effective potentials considered (cf. Eqs.~(\ref{eq:2bpot},
\ref{eq:3bpot})). Thus the scale $E_s$ defined in Eq.~\eqref{eq:E_s}
is the natural energy scale of the problem, 
and all energies are quoted in units of $E_s$. 
To embed our findings in a larger context, we show in Fig.~\ref{fig:overview} an Efimov plot for
$H_2 L$ trimer and $H_3 L$ tetramer states for 
the mass ratios $M/m = 133/6$ and $M/m = 7/6$.
All states shown are well within the universal window indicated by the shaded area in Fig.~\ref{fig:3body}.
The results for $M/m =87/7$ are not plotted because they would overlap with the results for $M/m = 133/6$. 
To increase visibility of the features of the plot, the axes are rescaled by taking the fourth root. The
variables $H$ and $\xi$ are introduced as $r_0/a = H\cos\xi$ and  $-\sqrt{\abs{E}/E_s} = H\sin\xi$
(cf.~Ref.~\cite{Hammer:2006ct}).\footnote{Due to the rescaling of $H$ to $H^{1/4}$ small deviations
  from the linearity of the dimer threshold are 
  strongly magnified.  Therefore we plot the energy differences to the $HL$-$H$-$H$ dimer-atom-atom threshold on the right half
  plane and  include an idealized dimer-atom-atom threshold.}
We only show the lowest trimer because we focus on the tetramer states attached 
to each trimer for two different mass ratios. We could confirm that, as already noted by Blume at al.~\cite{Blume2014}, the excited tetramer 
does exist in the unitary limit for $M/m = 133/6$ and does not for $M/m = 7/6$. They reported that for mass ratios 
less than 16, namely 12 and 8, they could only find an excited tetramer away from the unitary limit on the negative scattering length side. 
For $M/m = 7/6$ we did not find any 
excited tetramer, not even on the negative scattering length side.  This indicates that the value of $1/a$ where the excited 
tetramer crosses into the trimer threshold recedes from a point on the positive scattering length side of the Efimov plot for large mass ratios to 
points on the negative scattering length side for smaller mass ratios until it ceases to exist altogether. 
However, to prove this hypothesis, more data points are necessary. 
%inserted part
In a very recent study \cite{wangye2016}, the $^{23}$Na-$^{87}$Rb-system has been calculated. In this case
with a mass ratio of about $3.8$ no excited state was found, which agrees with our findings.
%end of inserted part

Comparing the curves for the two mass ratios, one can see that the $M/m = 7/6$ mixture has much shallower bound states, approaching 
the dimer threshold with an extremely small slope. This holds true even when the three-body potential strength is adjusted 
such that the trimer vanishes at the same value for $r_0/a$ for both mixtures. 
In addition, in both cases the tetramer and the trimer 
seem to vanish at the same point. 
%inserted part
This result was also corroborated in \cite{wangye2016} for the NaRb$_2$ trimer and the 
NaRb$_3$ tetramer.
%end of inserted part
In the next section, we show our results for this point with 
much higher resolution and discuss it in more detail.

%%%%%%%%%%%%%%%%%%%%%%%%%%%%%%%%%%%%%%%%%%%%%%%%%%%
%%%%%%%%%%%%%%%%%%%%%%%%%%%%%%%%%%%%%%%%%%%%%%%%%%

\section{Tetramer and Trimer Close to the Dimer Threshold}
\label{4and3closetothresh}
%%%%%%%%%%%%%%%%%%%%%%%%%%%%%%%%%%%%%%%%%%%%%%%%
\begin{figure}[htb]
\bigskip
% \centerline{\includegraphics*[width=12cm,angle=0]{figure133-6.pdf}}
\centerline{\includegraphics*[width=10.9cm,angle=0]{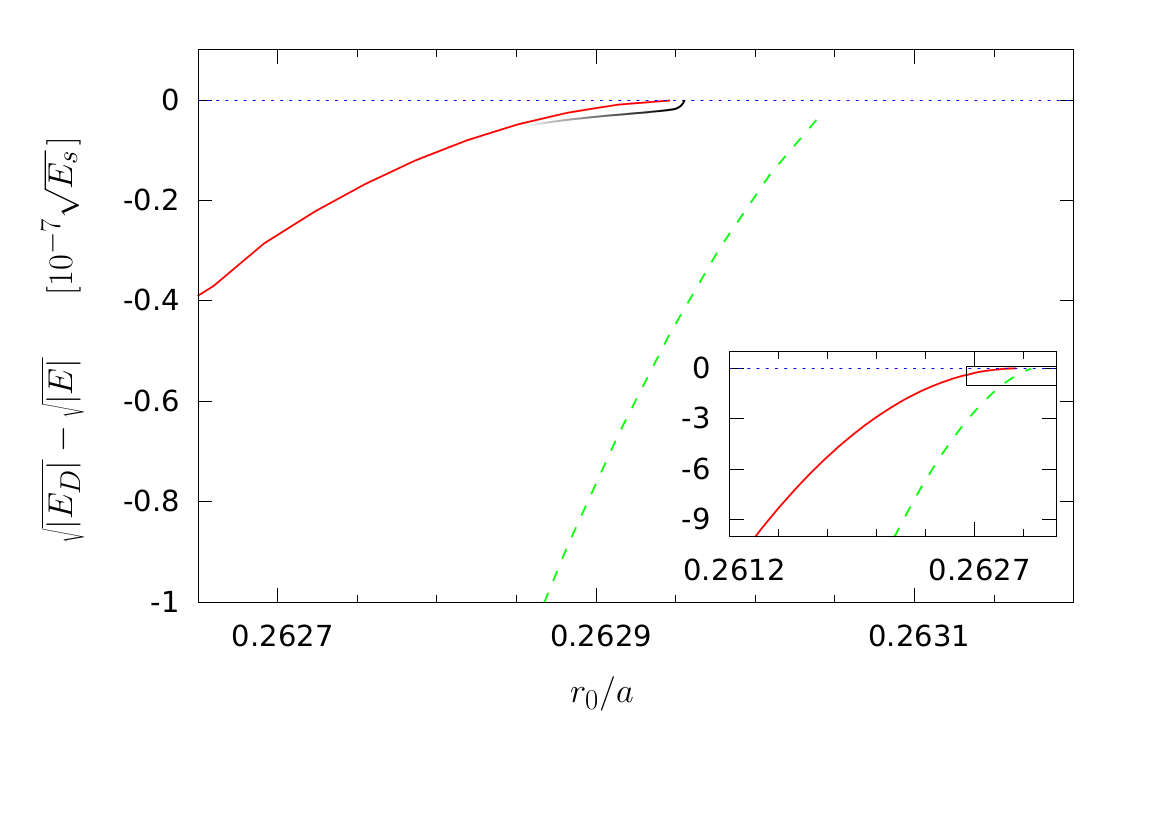}}
\medskip
 \caption{Binding energies of tetramer (dashed line) and trimer 
 (solid line) of the 133-133-6 system in the region close to 
 the $HL$ dimer threshold (dotted line). $E_D$ is the $HL$ dimer energy and
 $E_s$ is the natural energy scale defined in Eq.~\eqref{eq:E_s}. The inset 
 shows a larger region of $r_0/a$. We included our estimate (solid black line) of where the next excited Efimov state should appear 
  if the scaling factor were 20 (see discussion in Sec. \ref{resolutionestimate}). We did not find evidence for such a state.}
\label{fig:133-6}
\end{figure}
%%%%%%%%%%%%%%%%%%%%%%%%%%%%%%%%%%%%%%%%%%%%%%%%%

Next we focus on the tetramer spectrum close to the atom-dimer threshold
in the region where the trimer disappears. 
This region has not been studied in a full four-body approach 
for heteronuclear systems. 
%%%%%%%%%%%%%%%%%%%%%%%%%%%%%%%%%%%%%%%%%%%%%%%%
\begin{figure}[htb]
\bigskip
\begin{center}
%  \includegraphics[clip=true,trim=10pt 0pt 0pt 0pt]{./}
 % .: 0x0 pixel, 0dpi, 0.00x0.00 cm, bb=
\end{center}
\centerline{\includegraphics[width=\textwidth, trim=2.1cm 21.8cm 2.1cm 0.5cm]{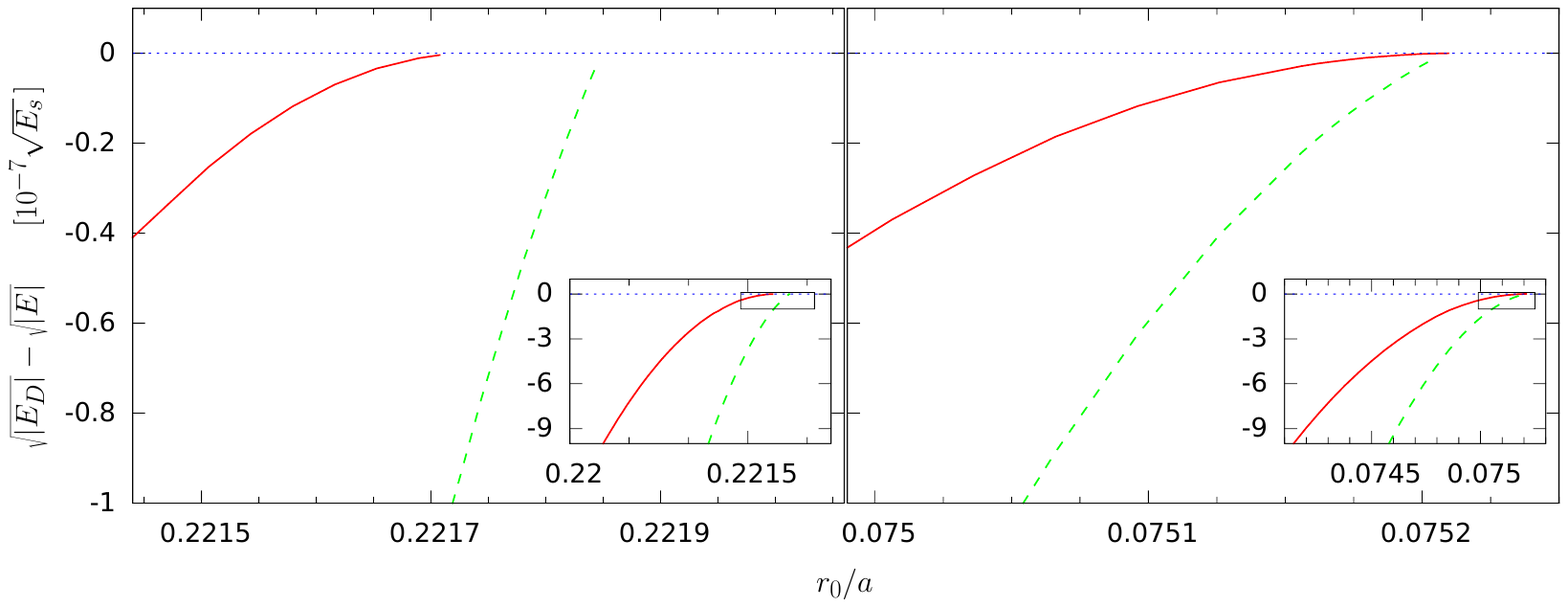}}
% \hspace{3.5ex}\includegraphics*[width=11.1cm,angle=0,trim=0 0.25cm 0 1.29cm]{figure87-7m.pdf}%
% \hspace{-8.7ex}\includegraphics*[width=9.1564cm,clip=true,trim=5.2cm 0.25cm 0cm 1.29cm]{figure7-6m.pdf}}
\medskip
 \caption{Binding energies of tetramer (dashed line) and trimer (solid line)
of the 87-87-7 system (left panel) and of the 7-7-6 system (right panel) in the region close to 
 the $HL$ dimer threshold (dotted line). 
 $E_D$ is the $HL$ dimer energy and
 $E_s$ is the natural energy scale defined in Eq.~\eqref{eq:E_s}. The inset 
 shows a larger region of $r_0/a$.}
\label{fig:87-7}
\end{figure}
%%%%%%%%%%%%%%%%%%%%%%%%%%%%%%%%%%%%%%%%%%%%%%%%%
The corresponding results are shown in 
Figs.~\ref{fig:133-6} and \ref{fig:87-7}, respectively. 
In all three cases ($M = 133, m = 6$), ($M=87, m=7$), and ($M=7, m=6$),
the tetramer disappears through the dimer threshold after the trimer has 
already disappeared.
However, the points in $r_0/a$ where the trimer and tetramer disappear approach 
each other as the mass ratio $M/m$ is decreased and the two states almost 
disappear at the same point in the ($M=7, m=6$) case. This is a qualitatively 
new behavior that is not seen in systems with identical bosons. 
It might suggest that for equal masses $M=m$, the two states disappear 
at the same value of $r_0/a$. We have further investigated this below.
If the mass ratio $M/m$ is even further decreased, we enter the 
regime of an inverted mass ratio where the scaling factor grows 
rapidly. In the current study we have not investigated this region. 

\begin{figure}[htb]
\bigskip
\centerline{\includegraphics*[width=12cm,angle=0]{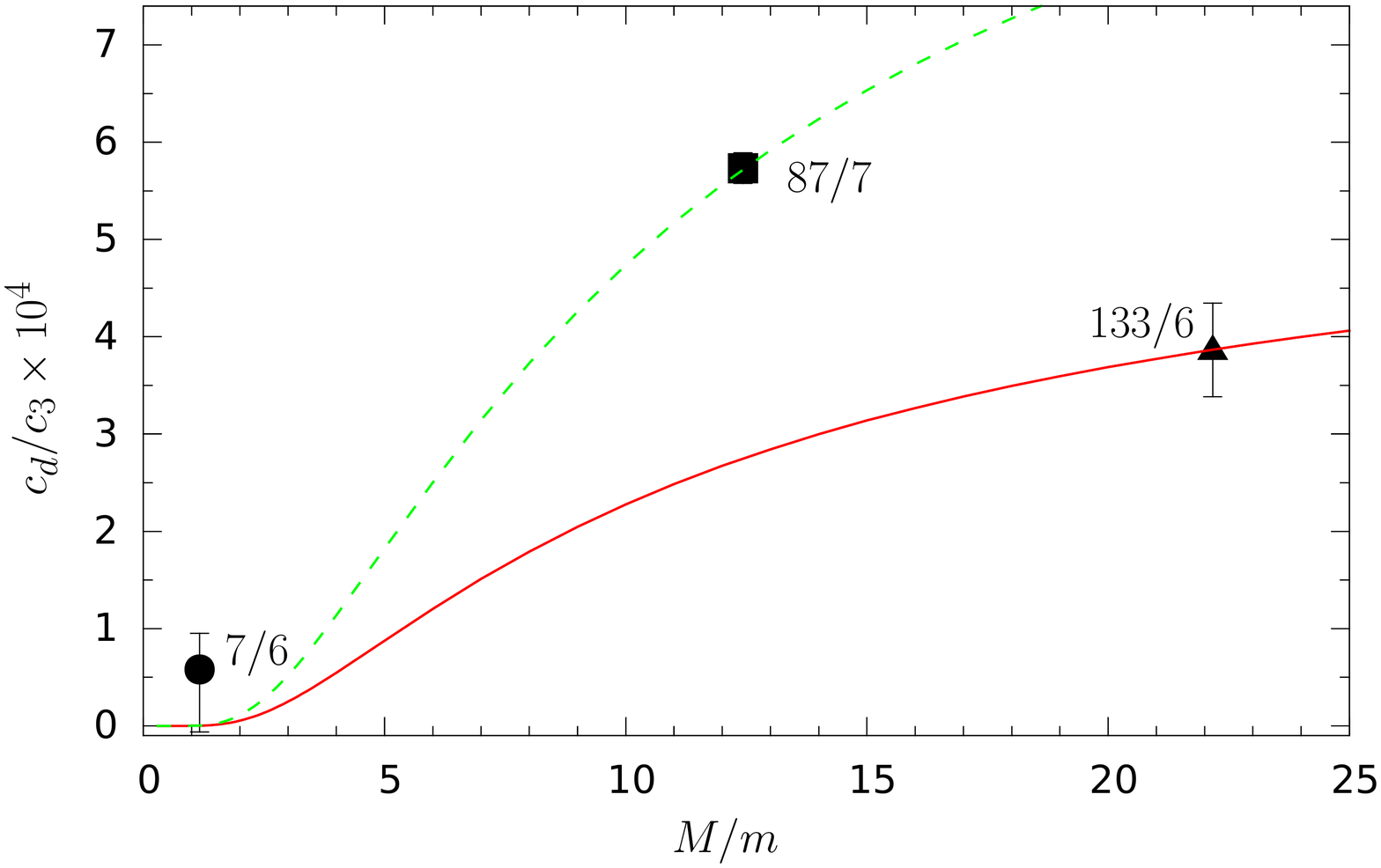}}
\medskip
\caption{Difference between trimer and tetramer
thresholds in units of $r_0/a$ as a function of the mass ratio
$M/m=7/6,\, 87/7,\, 133/6$ (circle, square, triangle). 
The dashed (solid) line shows effective field theory 
calculations for an effective three-body system of an $HL$ dimer and two 
heavy bosons $H$ of mass $M$ fitted to reproduce the point $M/m=87/7$
($M/m=133/6$). 
}
\label{fig:thresh}
\end{figure}

To quantify the behavior of the threshold crossings,
we have extracted the difference $c_d=c_4-c_3$ between the points
$c_4$, denoting the value of $r_0/a$ where the $H_3 L$ tetramer vanishes into
the $HL$+$H$+$H$-threshold, and $c_3$, the point where the $H_2 L$ trimer vanishes into 
the $HL$+$H$-threshold, from Figs.~\ref{fig:133-6} and \ref{fig:87-7}. 
We show this coefficient as a function of 
the mass ratio $M/m$ in Fig.~\ref{fig:thresh}.
%%%%%%%%%%%%%%%%%%%%%%%%%%%%%%%%%%%%%%%%%%%%%%%%
%%%%%%%%%%%%%%%%%%%%%%%%%%%%%%%%%%%%%%%%%%%%%%%%%
Our results for the mass ratios $M/m=7/6,\, 87/7,\, 133/6$
are shown by the circle, square, and triangle, respectively.
The errors in the extracted values are relatively large, but our 
calculations indicate that the relation is not linear. 

In order to understand this pattern, we have performed
effective field theory calculations for effective three-body systems
of an $HL$ dimer of mass $M+m$ and two $H$ bosons of mass $M$, where only the 
dimer and the bosons interact. Using the effective field theory 
formalism 
for unequal mass particles with resonant interaction discussed
in Ref.~\cite{Helfrich:2010yr}, we can calculate the coefficient 
$c_d$ up to one free parameter, namely the effective three-body parameter
of the $HL$-$H$-$H$ system. (See Appendix \ref{app:EFT} for a brief
discussion of this framework.) Therefore we can fit the effective three-body
parameter either to the point $M/m=87/7$ (dashed line) or $M/m=133/6$ (solid line). 
In each case 
the point $M/m=7/6$ is reproduced within error bands. 
A fit that reproduces all three points is not possible. 
This shows the limitations of the effective three-body calculation
which is strictly valid only near $c_d\approx 0$.
The effective three-body calculation shows
that $c_d$ vanishes for $M/m\rightarrow0$, and is very small but finite 
at $M=m$. 
We emphasize that this effective three-body approach
is not exact and should only be valid near the point where the trimer vanishes through
the $HL$ dimer threshold. 
An estimate of the $HL$-$H$ scattering length indicates that its value
is at least two orders of magnitude larger for the case $M/m=7/6$
compared to the other two mass ratios. Thus we expect the effective
three-body picture to work best for this case.

The tension between effective three-body 
and the GEM calculations could be due to two potential problems.
First, the GEM calculations could be not as accurate as we estimate,
possibly due to neglected higher partial wave contributions or 
hard to detect numerical pathologies. Second, some essential aspect of 
the system might not be present in the effective three-body description, 
which might also explain why different approximation schemes 
yield different results as discussed in the next section.

\section{Dimer-Atom-Atom Efimov States}

To discuss the effective Efimov states that were already mentioned in the introduction, 
three aspects are important. To begin with, the region where they can occur has to be 
established. For discussions on whether or not they can be seen in numerical calculations or even in experiments, 
an important issue is the predicted scaling factor. We will briefly present two competing approximation 
schemes that lead to very different predictions. To conclude, we will comment on 
the current state of the evidence for the effective Efimov states. 

\subsection{Motivation and Region of Possible Occurrence}

Near the dimer threshold, both trimer and tetramer become very weakly bound 
in comparison to the dimer. For this reason it is expected that at a certain point 
the system can be treated as an effective three-body problem. At the point 
were the trimer becomes unbound, the dimer-atom interaction is resonant, 
leading to the occurrence of the Efimov effect in the effective three-body system.
Therefore, the effective Efimov effect is expected to arise in a small region 
around the point where the trimer crosses the dimer threshold. This region 
is bounded by the tetramer ground state from below. As we have shown in 
the previous section, this region (the size of which can be expressed by $c_d/c_3$) proved to be extremely small in our 
calculations. 
The effective Efimov effect has already been studied in systems of four identical bosons 
by Deltuva \cite{deltuva2012dimeratomatom} and in heteronuclear systems by Wang et al. \cite{wanglaing2012}.

For four identical bosons, Deltuva found $c_d/c_3 < 1.2 \times 10^{-6}$, which compares well to 
our value for $^{7}$Li/$^{6}$Li, $c_d/c_3 = 5.8 \times 10^{-5}$. For equal masses, 
our effective three-body calculations yielded values between $10^{-8}$ and $10^{-7}$, 
depending on the three-body parameter (compare Fig.~\ref{fig:thresh} and the discussion in the previous section). 
Assuming that the system is away from the unitary limit and close to the dimer threshold the 
dimer-atom scattering length is 
much larger than the scattering length between identical bosons, which means that the 
system studied in \cite{deltuva2012dimeratomatom} is comparable
to our calculations. %Shifts due to the atom-atom interaction are of course possible.
Deltuva also did effective three-body dimer-atom-atom calculations and showed that very close 
to the dimer-atom-atom threshold this approximation is valid. 
His estimated scaling factor  for the effective Efimov states is 
$5\times 10^{5}$, which agrees with \cite{Braaten:2004rn} and our discussion in the next section.

In a paper from 2012, Wang et al. \cite{wanglaing2012} used the Born-Oppenheimer 
approximation in heteronuclear systems with very high mass ratios ($M/m = 30$ and $50$) to 
calculate recombination features. They calculated trimer ($H_2L$), tetramer ($H_3L$) 
and also $H_2(HL)$ states as effective two- and three-body problems. They also used 
correlated Gaussian results to check their calculations where possible. 
Moreover, they provide a qualitative sketch of the heteronuclear Efimov spectrum. 
We note, however, that this sketch is somewhat misleading concerning the window where 
dimer-atom-atom Efimov states could form, as its size is exaggerated in light of 
our results.  Comparing their results to ours, 
several differences can be noted.

From Fig.~2 in~\cite{wanglaing2012} we extracted their $c_d/c_3$ for $M/m=30$, which
is roughly $0.03$, an order of magnitude larger than 
for our calculations (extrapolations via the effective three-body calculations 
to $M/m = 30$ suggest something of order $10^{-3}$).
In Ref.~\cite{wanglaing2012}, separable potentials of Yamaguchi type between the $HL$ pairs
were used \cite{wang2017private}.
So there are three obvious possible reasons for the
discrepancy, a)  $c_d/c_3$ is not universal and in fact depends on the $HL$ potential, 
b) the BO approximation is problematic in this region, as may be suggested 
by the fact that the scaling factor between consecutive Efimov states is 5.5 in their calculations, 
while the universal value is 3.96~\cite{Braaten:2004rn},
and c) the effective three-body calculations we used are not valid for higher mass ratios 
because of the larger $c_d/c_3$ value, and therefore the extrapolation to $M/m=30$ is not sound.

To rule out a), one would have to use a different potential and redo the calculations. 
Moreover, it is unclear whether the calculations of Wang et al. \cite{wanglaing2012} are conducted closer to the unitary 
limit than ours, but since they state that $a_{HL}\gg r_0$ is not well satisfied in their 
calculations, we suppose $r_0/a_{HL}$ is of comparable size for both their calculations and ours (we have 
$r_0/a_{HL} \approx 1/4$ for the ($87/7$) and ($133/6$) systems, and $r_0/a_{HL} \approx 1/13$ for the ($7/6$) system). 
In addition, because $c_d/c_3$ is very similar for the excited and ground state trimer in their calculation, 
and of roughly the same order of magnitude for different values of $r_0/a_{HL}$ in our calculations, 
a strong dependence on $r_0/a_{HL}$ is not likely.

Following the argument b), one might also add that our effective three-body 
calculation did not agree too well with our full four-body calculation, as shown in the 
previous section. (Note that the order of magnitude of $c_d/c_3$ is a fit parameter in 
our STM treatment, so it agrees by construction.) This might indicate that approximations that 
reduce the problem to an effective three-body system are generally insufficient to 
accurately describe the behavior near the dimer threshold. 

Aspect c) should not play a huge role because the atom-dimer scattering length $a_{(HL)H}$
is estimated to be larger than $10^{7}r_0$ for $M/m =133/6$ and does not change much between 
$M/m = 87/7 \approx 12$ and $M/m=133/6 \approx 22$, so extrapolating to $M/m = 30$ 
should yield reasonable accuracy.

\subsection{The Scaling Factor}

The dimer-atom-atom Efimov 
effect is conjectured to arise when the four-body system of $H_3L$ can be 
treated as an effective three-body system that has an effective resonant 
interaction. This is expected to be the case in the vicinity where the $H_2L$ three-body 
state vanishes into the dimer threshold. 
There, Braaten and Hammer \cite{Braaten:2004rn}
argued, we can regard the $HL$ dimer as much more strongly bound than the trimer, 
allowing us to treat the trimer as an effective two-body system $(HL)H$ with an 
effective $(HL)-H$ scattering length $a_{(HL)H}$ which can be estimated by the 
inverse square root of the binding energy of the $H_2L$ trimer. Then the $H_3L$ tetramer 
can be treated as an effective three-body system $H_2(HL)$ which is also governed by 
the dimer-atom scattering length $a_{(HL)H}$. 
For large $a_{(HL)H}$ this leads to the emergence of the 
Efimov effect. The scaling 
factor expected in this case is roughly 2000, because there are only two interacting pairs 
$(HL)-H$, and there are two ``light'' particles $H$ and one that is slightly heavier $(HL)$. 
The scenario with the smallest scaling factor is $m/M \rightarrow 0$, in which case it is 
1986.1 \cite{Braaten:2004rn}. This is the picture we have been using in our STM calculations shown 
in Fig.~\ref{fig:thresh}. Because of the huge scaling factor, we did not expect to be able 
to find these Efimov states with the accuracy we have reached. 

However, in the Born-Oppenheimer approximation picture that Wang et al. \cite{wanglaing2012}
used, the light atom $L$ is seen as mediating the interaction between the heavy 
bosons $H$, leading to an effective three-body problem of three $H$'s, with an 
effective scattering length $a^*_{HH}$. This effective scattering length diverges 
for the values of $a_{HL}$ where an Efimov trimer becomes unbound, so it shows the 
same behavior as $a_{(HL)H}$. But, since in this picture the $L$ is absorbed in the 
effective interaction, a symmetric system of three resonantly interacting $H$'s emerges. 
The Efimov effect for this system, which is the well known three identical boson case, yields 
a much smaller scaling factor of $e^{\pi/s_0} \approx 22.7$ \cite{Braaten:2004rn}. 

Using the Born-Oppenheimer approximation, Wang et al. \cite{wanglaing2012} find 
somewhat tenuous evidence for an excited tetramer separated from the lower tetramer by 
a scaling factor of $14.3$ on the positive $a^*_{HH}$ side and $19.3$ on the 
negative side. This is in disagreement with the expectation from the EFT 
approach outlined above. 

The main difference between the two approaches lies in the 
way the separation of scales is conducted. In the first approach, it is assumed 
that the $HL$ dimer is much more deeply bound than the $H_2L$ and $H_3L$ states, allowing 
a separation of energy scales and associated length scales. This 
assumption is found to be valid, because $\abs{E_D}/\abs{E_3^0-E_D} > 300$ for the whole 
region shown in Figs.~\ref{fig:133-6}\ and \ref{fig:87-7}. On the other hand, 
in the BO approximation it is assumed that the time scale of the motion of the light atom is much 
shorter than that of the heavy atoms, allowing us to solve the dynamics of the light atom 
for fixed heavy atoms and then solving the heavy atoms system independently. This 
assumes that the light atom interacts with all the heavy atoms while they are fixed at 
some points, thus generating an effective interaction. 

If however the light atom is very tightly bound to one heavy atom, as is the case here, 
its motion around the heavy atom might be on a shorter timescale than the motion of 
the heavy atom, but its motion towards the other heavy atoms is coupled to the motion 
of the heavy atom, which means it is on the same timescale and therefore not separable. 
%inserted part
So care would have to be taken to ensure that the light atom only induces an 
interaction of the heavy atom it is localised on with the other two heavy atoms, 
but not between the other two heavy atoms.
%end of inserted part

This means that in our understanding, both effective treatments are mutually exclusive 
and have to be applied to different systems. If we regard the two approximations as 
two ends of a spectrum, the BO approximation 
%inserted part
as used in \cite{wanglaing2012}
%end of inserted part
would apply to the case where the dimer 
and the trimer are both very weakly bound, and the EFT approximation would apply to the 
case where the dimer is infinitely strongly bound. Between these two points, 
we conjecture there is a transition between the three identical boson case with its relatively small scaling factor 
which then becomes 
larger and larger until it reaches the limit of two light bosons with one heavy boson. 
For the case we are studying here, this could mean that the scaling factor is slightly smaller 
than we would expect from our STM calculation.
This however can only be explored using a full four-body calculation that is accurate enough 
to resolve these effective Efimov states.

\subsection{Resolution Estimate}
\label{resolutionestimate}
To estimate whether it would be possible to see the effective Efimov states 
in our calculations, we made some further assumptions. First, we assumed 
that the universal tetramer can be interpreted at the same time as the 
lowest state of the effective Efimov states. We do not expect this state 
to be fully universal, so the scaling factor between this state and the next 
may be smaller or larger, but we expect it to be of the same order 
of magnitude. This assumption was apparently also made by Wang et al. \cite{wanglaing2012}, 
although they did not mention it explicitly.

The second assumption is that the shape of the first excited effective Efimov state would 
resemble the shape of the universal tetramer near the threshold. Then we could scale the 
tetramer in Figs.~\ref{fig:133-6} and \ref{fig:87-7} using different scaling factors to assess 
whether this state could be found with our accuracy. An example for a scaling factor of 20 is 
shown in Fig.~\ref{fig:133-6}.

Our four-body calculations presented here are by far not 
accurate enough to find effective Efimov states with a scaling factor of 2000. But our accuracy might 
just be sufficient to detect effective Efimov states if the scaling factor was indeed approximately 20
as suggested by Wang et al. \cite{wanglaing2012}. We did not find any states, but since they would 
be just at the limit of our resolution, this is no conclusive evidence of their 
non-existence.
%%%%%%%%%%%%%%%%%%%%%%%%%%%%%%%%%%%%%%%%%%%%%%%
\section{Summary and Outlook}
\label{sec:outlook}
%%%%%%%%%%%%%%%%%%%%%%%%%%%%%%%%%%%%%%%%%%%%%%%
In this paper, we have presented a detailed study of the
bound state spectrum of heteronuclear four-body systems
of three heavy bosons $H$ and a light atom $L$ with resonant
interspecies interaction in the region of positive
scattering lengths. The interaction between the heavy bosons
was assumed to be negligible.
To obtain the energies of the $H_2L$ trimer
and $H_3 L$ tetramer systems,
we employed the Gaussian expansion method (GEM) \cite{hiyamakino2003}. 

The general structure of the spectrum for different mass ratios was
summarized in the Efimov plot of
Fig.~\ref{fig:overview}, a notable feature being that there is no excited Efimov tetramer at all
for the mass ratio $M/m = 7/6$. As the mass ratio increases, the excited state tetramer appears
on the negative scattering length side of the Efimov plot. For $M/m = 133/6$ it exists also
in the unitary limit and disappears through the trimer-atom threshold on the positive
scattering length side of the Efimov plot. Our results confirm and extend the findings
of Blume and Yan~\cite{Blume2014},  who conducted calculations down to $M/m = 8/1$. 

Next we focused on the dimer-atom(-atom) threshold region. We found that the value of the
scattering length where the trimer vanishes into the threshold and the point where the tetramer
does the same are extremely close to each other. We have investigated the dependence
of the distance between these points, $c_d$, on the mass ratio
$M/m$ and presented a detailed discussion of the sources of error in our calculation,
such as a possible dependence of the results on the potential shape and missing higher partial
wave contributions. We have estimated the size of errors from the latter to not be larger 
than $10^{-8} \sqrt{E_s}$ near the dimer-atom-atom threshold.
% This is especially relevant for the investigation of the
% distance between the two crossing points, $c_d$, on $M/m$, which requires high accuracy.
The correlation between $c_d$ and $M/m$ does not appear to be linear, and the results from the four-body calculation 
could be reconciled only partly with effective three-body results in the framework of a
STM-like equation derived from effective field theory~\cite{Helfrich:2010yr}.
We conclude that either an important aspect of the four-body system is missing in the effective
three-body calculation close to the dimer-atom-atom threshold, or the errors of our 
four-body results are underestimated. The resolution of this question requires further work,
such as repeating the calculations for different potential shapes, to test the influence on $c_d$
and the general shape of the tetramer near the threshold.
In order to rule out finite range effects a study of higher 
excited trimers and the tetramer resonances belonging to them 
would be useful.

In the last part, we discussed in depth the possibility of finding effective dimer-atom-atom 
Efimov states near the dimer-atom-atom threshold. We did not observe such states
and estimated it is unlikely to find them without a major improvement of the accuracy
in our calculation. Wang and collaborators~\cite{wanglaing2012} have previously found such states, 
albeit only in effective three-body calculations within the Born-Oppenheimer approximation.
We have compared our results to the results of ~\cite{wanglaing2012} and highlighted some important
differences between the two approaches. The key question in reconciling the results involves the
scaling factor for the effective Efimov states. We expect a large scaling factor of order 2000
since only the $HL$ dimer and the $H$ atoms of the effective three-body system are interacting resonantly.
We estimated that for this scaling factor an observation of the effective Efimov states in our calculation
is very unlikely. In the Born-Oppenheimer
picture, however, the light $L$ atom mediates an interaction between all three $H$ atoms such that
there are effectively three resonant pair interactions and  the
scaling factor should be much smaller and close to 22.7.
As we discussed in detail, it is not obvious that the
Born-Oppenheimer picture is still applicable near the dimer-atom-atom threshold.
Settling the question of the effective dimer-atom-atom Efimov states in heteronuclear systems requires a significant
increase in the accuracy of our four-body calculations and a coordinated study of the problem in
both approaches.  

\begin{acknowledgments}
We acknowledge useful discussions with Doerte Blume on heteronuclear few-body systems
and practical units,
and with Artem Volosniev on the limitations of the Born-Oppenheimer approximation. 
We also thank Doerte Blume for providing data from her figures from \cite{Blume2014}.
This research was supported in part by the Deutsche Forschungsgemeinschaft through SFB 1245
and by the German Federal Ministry of Education and Research under contract 05P15RDFN1.
C.H.S. thanks RIKEN for support under 
the IPA program and the Studienstiftung des deutschen Volkes for support under the RIKEN program. 
Some of the calculations were carried out at supercomputing facilities of the YITP (Kyoto University) and
of Nagoya University.
\end{acknowledgments}

\appendix
\section{Effective field theory framework}
\label{app:EFT}
An effective field theory for heteronuclear three-body systems
consisting of an atom of mass $m_1$ and 
two identical bosonic atoms of mass $m_2$
was developed by Helfrich et al.~\cite{Helfrich:2010yr}.
This effective field theory leads to an analog of the
Skorniakov-Ter-Martirosian (STM) equation~\cite{STM:1957}
for the unequal mass case. Similar equations also arise in
the effective field theory description of 
two-neutron halo nuclei in nuclear physics~\cite{Canham:2008jd}.
Here we focus on the case of zero-range interactions between the unlike
atoms characterized by a scattering length $a_{12}$
and no interaction between the like atoms ($a_{22}\equiv 0$).

The trimer binding energies $B$ in this framework are given by the non-trivial
solutions of the integral equation:
%-------------------
\begin{equation}
\label{boundstateeq}
\chi(p;B)=\frac{m_1}{2\pi\mu_{12}}\int_0^{\Lambda}
dq\,\frac{q}{p}\,\ln\left(\frac{p^2+q^2+2pq\mu_{12}/m_1 +2\mu_{12} B}
{p^2+q^2-2pq\mu_{12}/m_1 +2\mu_{12} B}\right)\,\frac{\chi(q;B)}
{-\frac{1}{a_{12}}+\sqrt{2\mu_{12}\left(B+\frac{q^2}{2\mu_{(12)2}}\right)}}\, ,
\end{equation}
%---------------------
where $\mu_{12}=m_1 m_2/(m_1+m_2)$ and $\mu_{(12)2}=(m_1+m_2)m_2/(m_1+2m_2)$
are the reduced
masses of the $12$ and $(12)2$ systems. The cutoff $\Lambda$
plays the role of the three-body parameter and can be used to fix
the energy $B$ of any three-body state in the system.\footnote{Note
that in Eq.~(\ref{boundstateeq}), the log-periodicity of the three-body
force term in the effective field theory 
was used to move the dependence on the three-body parameter
to the upper limit of the integral (see Ref.~\cite{Hammer:2000nf} for
more details).} All other three-body energies can then be predicted.
Alternatively, the energies can be predicted in units of $\Lambda^2$.
The wave function of the trimers can be obtained from the function
$\chi(p;B)$ but will not be required here.

\end{document}